UDC 004.5: 004.9: 37

# Developing Educational Computer Animation Based on Human Personality Types

Sajid Musa[1], Rushan Ziatdinov[1,*], Omer Faruk Sozcu[1], Carol Griffiths[2]

[1]*Department of Computer and Instructional Technologies, Fatih University, 34500 Buyukcekmece, Istanbul, Turkey*
*sajidmusa004@gmail.com, rushanziatdinov@gmail.com, ofsozcu@fatih.edu.tr*
[2]*Department of Foreign Language Education, Fatih University, 34500 Buyukcekmece, Istanbul, Turkey,*
*carolgriffiths5@gmail.com*

**ABSTRACT**

Computer animation in the past decade has become one of the most noticeable features of technology-based learning environments. By its definition, it refers to simulated motion pictures showing movement of drawn objects, and is often defined as the art in movement. Its educational application known as educational computer animation is considered to be one of the most elegant ways for preparing materials for teaching, and its importance in assisting learners to process, understand and remember information efficiently has vastly grown since the advent of powerful graphics-oriented computers era. Based on theories and facts of psychology, colour science, computer animation, geometric modelling and technical aesthetics, this study intends to establish an inter-disciplinary area of research towards a greater educational effectiveness. With today's high educational demands as well as the lack of time provided for certain courses, classical educational methods have shown deficiencies in keeping up with the drastic changes observed in the digital era. Generally speaking, without taking into account various significant factors as, for instance, gender, age, level of interest and memory level, educational animations may turn out to be insufficient for learners or fail to meet their needs. Though, we have noticed that the applications of animation for education have been given only inadequate attention, and students' personality types of temperaments (sanguine, choleric, melancholic, phlegmatic, etc.) have never been taken into account. We suggest there is an interesting relationship here, and propose essential factors in creating educational animations based on students' personality types. Particularly, we study how information in computer animation may be presented in a more preferable way based on font types and their families, colours and colour schemes, emphasizing texts, shapes of characters designed by planar quadratic Bernstein-Bézier curves. The study has found out that both choleric-melancholic and phlegmatic-sanguine gained the lowest and the highest percentages in selection of different colour groups as cool, warm, and achromatic. We have experimentally confirmed the theory of Nabiyev & Ziatdinov (2014) which reports that planar quadratic Bernstein-Bézier curves with monotonic curvature function may be not aesthetic. Finally, based on the survey results, we have clarified how school students understand the fundamental principles of computer animation. We look forward that this study is likely to have wide benefits in the field of education. Developing educational materials with the aid of obtained empirical results, while considering the personality types of students' temperament, seems to be a promising avenue to improve, enrich and deepen the learning process in order to achieve its maximum effectiveness.

# Introduction

Recent lectures and talks by this manuscript's authors (Musa et al., 2013) in the Catholic University of Ružomberok in Slovakia have shown that there is a certain interest in educational computer animation from various scientific and educational schools of Europe. Computer animation which is one of the highly-recommended ways of explaining models and processes in natural science education has spread its influence on learning and instruction in areas like economy, industry, medical tourism and their educational aspects. Before we discuss computer animation itself, we review its history.

It has been twenty years since animation became the most notable feature of the technology-based learning environment (Dundar, 1993). Mayer and Moreno (2002) defined animation as a form of pictorial presentation, referring to computer-generated moving pictures showing associations among drawn figures. Motion, picture and simulation correspond to this idea. Videos and illustrations are motion pictures portraying movement of real objects.

Verbal forms of teaching have been augmented by pictorial forms of teaching (Lowe, 2004; Lasseter, 1987; Pailliotet & Mosenthal, 2000). Although it is undeniable that verbal modes of presentation have long reigned supreme in education, the addition of visual forms of presentation have enhanced students' understanding (Mayer, 1999; Sweller, 1999). As a matter of fact, animation or graphic illustration is preferred to verbal or numerical presentation by most university students when dealing with dynamic subject matter (Lowe, 2004).

Nevertheless, Lowe (2004), Lasseter (1987), and Pailliotet et al. (2000) note that the creation of multimedia instructional environments – holding potential for enhancing learner's way of learning – has created much debate. It is evident that animation presentations are less useful than was expected. In addition, there is inadequate knowledge about how to create animation in order to aid learning (Plötzner & Lowe, 2004) and some create it for the sole purpose of gaining aesthetic attraction. According to Lowe (2004), some who work in the entertainment industry tend to create characters just for entertainment, rather than using it as a bridge which would help to build coherent understanding using their work.

Several cases have shown that animation can even hold back rather than improve learning (Campbell et al., 2005) depending on how it is used (Mayer & Moreno, 2002). Besides, cognitive connection can be lost since animation imposes greater cognitive processing demands compared to static visuals since the information drastically changes (Hasler, 2007).

According to Mayer (2005), *"The current emphasis on ways of improving animations implicitly assumes a bottom-up model animation comprehension… Comprehension is primarily a process of encoding the information in the external display, so that improving that display necessarily improves understanding".* The role of animation in multimedia learning examined by Mayer and Moreno (2000) showed a cognitive theory of multimedia learning. They were able to name seven principles for the application of animation in multimedia instruction. To name one of them, according to the multimedia principle, students absorb more when both narration and animation are used rather than using just one or the other. When these two are presented together, learners can easily create mental connections between corresponding words and pictures. According to the coherence principle, learners learn productively both from animation and narration especially when unnecessary words, sounds (even music) and videos are not present. The reason behind this is the trouble the learners experience when creating mental connections due to fewer cognitive resources between relevant portions of the narration and animations (Plötzner & Lowe, 2004).

Educational computer animation also plays an important role in assisting language teaching and learning (Bikchentaeva & Ziatdinov, 2012; Musa et al., 2013). The effect of learner controlled progress was examined by Hasler (2007) regarding educational animation on instructional effectiveness. Referring to her findings, to teach the determinants of day and night to primary school students, three audio-visual computer animations and narration-based presentations were used. Moreover, Hasler (2007) noted that two of the groups displayed higher test performance compared with the other, based on the results of the experiment.

Taking into account the limitations of the studies discussed above, we raise the research question, "what is the connection between human personality types and their preferences in selecting the fundamental elements (variables) and understanding the principles of computer animation?" All these are important for better understanding of human psychology in order to find yet unknown methods for influencing human's visual perception, as well as their use for educational purposes.

**Main results**

In this paper, we discuss theoretical aspects of creating educational computer animations based on psychological characteristics of human personality types. Particularly, we study how information in computer animation may be presented in more efficient ways based on font types and their families, colours and colour schemes, emphasizing texts, shapes of planar quadratic Bernstein-Bézier curves. We analysed how tested students understand the fundamental principles of traditional computer animation, and question whether these principles should or should not be followed for developing educational computer animations for use in high schools of Turkey, where computer science or informatics classes may not even exist.

Our work has the following novelties:

- For the first time in computer-assisted learning we have proposed a way for developing educational computer animations based on fundamental personality types of human temperament;
- We have experimentally analysed how fundamental principles of traditional computer animation are understood by school students;
- We examined the relationship between personality type and preference for popular fonts and font families, text emphasis, colour and shape, which are known as the fundamental elements of a computer animation.

**Temperament based learning**

In this section, we discuss the work which has been done in this field. The influence of temperament mainly within the context of the family and school on the development of school-age children was examined by McClowry (1992). In line with this, she presented examples of how nurses may use temperament theory when advising caretakers. Temperament theory (McClowry, 1992) serves as a guide for the nurses in assessing children's behaviour.

Graham (1995) offered recommendations in carrying out temperament-based intervention for parents alone or for both the parents and teachers. She concluded that "acknowledging temperament-based guidance would show that each child has a particular behavioural style that contributes to his or her development and to the social environment".

Jong et al. (2013) stated that the posture and muscle loading of the body is significantly affected by various different interactive gaming controllers. It has been argued that the period of exposure to the interactive gaming controllers affects the success in using the game for the purposes of learning. Jong et al. (2013) aimed to explore the different behavioural responses based on the different temperaments regarding mathematical game play by comparing the touch-based and gesture-based interactive devices among 119 kindergarten participants. The results indicated that the touch-based interaction (TBI) group compared to the gesture-based interaction (GBI) group performed better with respect to numerical counting in both games. It also showed that with all the dimensions of temperaments, persistence was the only one which had positive correlation to TBI. In other words, TBI was more favoured over GBI for kindergarten children. Jong et al. (2013) added that more emphasis on TBI would be a great move for e-learning designers.

Recently, Ali (2007, 2008) discussed the development of software tools for online teaching that enables a tutor to control the psychological state of students in the process of testing, and developed the mathematical model of a dual system of testing consisting of three parts: lectures, tests, and teaching program. This shows that a temperament-based teaching approach finds some applications in e-learning.

Li et al. (2007), adopting the Eysenck Personality Questionnaire (EPQ) (Eysenck, 1958), conducted a survey with 1620 student participants. Students were categorized based on their year level respectively: primary 5, junior secondary 2 and senior secondary 2. They found a significant correlation between personality types according to the EPQ and maths achievement. Both psychological factors and mathematics achievement have become a major area of study, and the relationship between them is strong. Although there are some scholars who think that personality does not have a great impact on academic achievement of students (Gao, 1994), others believe that temperament is an important factor (e.g. Li, 1994, 1996, Qiaorongning, 2003, Li et al., 2004)  Li et al. (2007) defined *superior* and *inferior temperaments* in terms of achievement in maths. Superior temperament-based students are those who have sanguine, sanguine-phlegmatic and phlegmatic types of temperament; they are the ones who benefit from learning mathematics. Benefitting refers to how students learn mathematics easily.  Students who are choleric, choleric-melancholic and melancholic which are considered inferior temperaments do not benefit from learning mathematics. Li et al. (2007) stated that in mathematics education, one should recognize students' temperament differences which indeed affect learning mathematics.

**The four temperament types**

In this section, we discuss the four temperament types shortly. We compare temperament types by their general descriptions, wonderful characteristics and unpleasant traits.

Direct discussion of the temperament types without looking at their roots would be a great mistake. Thus we would like to shortly describe how today's temperament types came to be. Technically, *"temperament"* from the Latin word *"temperare"* meaning *"to mix"* has a long history, from the ancient time of Greek physician Hippocrates

*3*

(460-370). Galen (AD 131-200) pioneered the first typology of temperament and gave the names "sanguine", "choleric", "melancholic" and "phlegmatic". These four were the temperamental categories named after bodily humours. We can easily see and understand these four temperaments by looking at Table 1 and their emotional representation shown in Fig. 1.

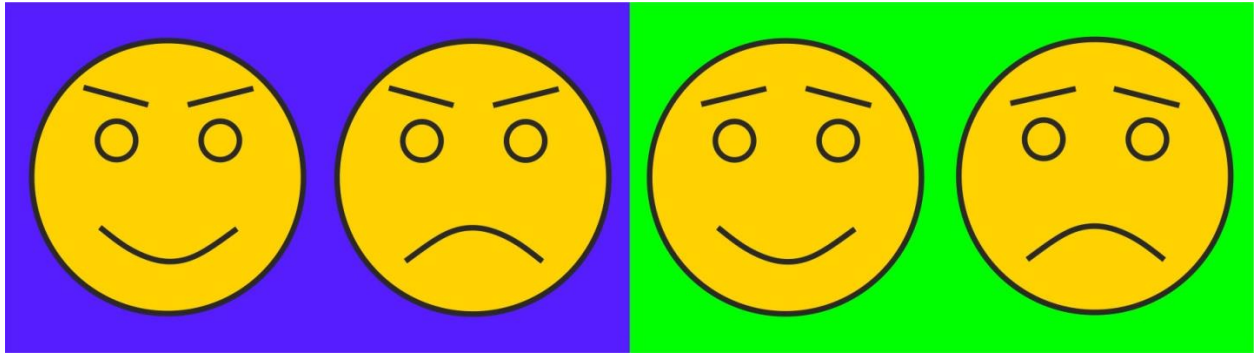

**Fig. 1.** Emoticon representation of the four temperament types. From left to right: sanguine and choleric (emotionally unstable), phlegmatic and melancholic (emotionally stable).

Hans Eysenck initially theorized personality as two dichotomies; extraversion/introversion and neuroticism/stability.

**Extraversion/Introversion**

Assertive, outgoing, sociable and talkative are only a few characteristics of extraversion. Based on Eysenck's arousal theory of extraversion, performance failure is due to one's inability to meet the ideal (optimal) level of cortical arousal. The person who does less or more than the desired optimal level ends up with under-performance. Brain waves, skin conductance or even sweating is used to measure arousal. Performance is low when the levels of arousal are very low and very high, however, performance is maximized at a more optimal mid-level of arousal. Therefore, according to Eysenck's theory, extraverts are chronically under-aroused and bored. In line with this, for extraverts to bring out the optimal level of performance, external stimulation is needed. Conversely, introverts are in need of peace and calmness to bring out the optimal level of performance. They are chronically over-aroused and nervous: that is why they need to be content from inside.

**Neuroticism/Stability**

Depression and anxiety are high levels of negative affect which define neuroticism or emotionality. Based on Eysenck's theory, the activation threshold of the visceral part of the brain or the sympathetic nervous system is neuroticism. The part of the brain tasked for the fight-or-flight response in case of danger is the sympathetic nervous system. Blood pressure, cold hands, heart rate, muscular tension and sweating measure the activation. A negative effect of fight-or-flight is experienced by those who have low activation thresholds in the case of minor stressors. In addition, since they are unable to inhibit their emotional reactions, they are easily nervous or saddened. These people are the neurotic ones. In the case of those who have high activation thresholds, they are the ones who have good emotional control. They are emotionally stable people. Nevertheless, when faced with very major stressors, they experience negative effects.

These two initial concepts, define four quadrants; *stable extraverts, unstable extraverts, stable introverts,* and *unstable introverts*. Human temperaments fall into these quadrants.

Moving on to the main discussion, there are four temperament types, namely: sanguine, choleric, melancholic and phlegmatic (Tab. 1). Each of the four types of humour is matched to a different personality type.

**Table 1.** Short summary of the four temperament types.

| Temperament | General Description | Wonderful Characteristics | Unpleasant Traits |
|---|---|---|---|
| Sanguine | easily-influenced; | compassionate; | flirt; |

|  | moody; optimistic; rarely internalizes. | friendly; humane; strong-willed; virtuous. | inner self-conflict; jealous; poor-decision maker; self-complacency; vain |
| --- | --- | --- | --- |
| Choleric | dreamer; dominator; enthusiastic; perfectionist; success-hungry. | clever; diligent; great rhetorical skills; leader. | hard-headed; prideful; talkative; belittle others; self-praised. |
| Melancholic | low rhetorical skills; strong principles; passive; self-reflection; serious; silent; thinker. | genuine; great adviser; loves solitude; successful; trustworthy; willing to sacrifice for others; | despairing; unforgiving; pessimistic; self-pity. |
| Phlegmatic | careless; feels emptiness; moves at a slow-pace. | contented in life; maintains composure; persevering; practical judgments; not easily offended. | lazy; no aspirations; takes things for granted. |

## Methods

### Research design

The experimental part of this research was conducted in Yaşar Acar Science High School, Beylikdüzü, Istanbul, Turkey. The participants were 240 students: 111 male and 129 female students between 14–17 years old [mean (M) = 15.53, standard deviation (SD) = 1.31]. The data was collected through a survey which consists of three parts. The first part is demographic information, the second part is the Eysenck Personality Questionnaire (EPQ), and the third part includes principles of animation.

### Data Collection

Surveys were applied to 252 volunteer students in December, 2013, and were administered to students in one class period and it took approximately 20 minutes. After data were collected, it was determined that 12 students' surveys were incomplete or answered without reading carefully. Thus, these data were excluded from the study.

### Participants

The participants' demographic information is given in Table 2 below:

| **Table 2.** Demographic analysis of data. | | | |
| --- | --- | --- | --- |
|  |  | Frequency | Percent |
| Gender | Female | 129 | 53.75 |
|  | Male | 111 | 46.25 |
| Age | 14 | 39 | 16.25 |
|  | 15 | 86 | 35.83 |
|  | 16 | 65 | 27.09 |
|  | 17 | 50 | 20.83 |
| Class | 9 | 102 | 42.50 |
|  | 10 | 56 | 23.33 |
|              |                    | 11  | 82     | 34.17 |
| Having PC    | Yes                | 206 | 85.83  |
| or tablet    | No                 | 34  | 14.17  |
|              | Baskerville Old Face | 12 | 5.00 |
|              | Tahoma             | 21  | 8.75   |
|              | Cambria            | 30  | 12.50  |
|              | Garamond           | 16  | 6.67   |
| Best Reading | Batang             | 3   | 1.25   |
| Font         | Bookman Old Style  | 21  | 8.75   |
|              | Times New Roman    | 69  | 28.75  |
|              | Verdana            | 15  | 6.25   |
|              | Calibri            | 35  | 14.58  |
|              | Gill Sans MT       | 18  | 7.50   |
|              | Bold               | 150 | 62.50  |
| Font Style   | Italic             | 43  | 17.91  |
|              | Highlighted        | 34  | 14.17  |
|              | Underline          | 13  | 5.42   |
|              | Total              | 240 | 100.00 |

## Analysis of data

**Purpose and frequency of computer use**

Table 3 below shows the results gathered partly from the survey regarding how frequently the students use computers and for what purpose. Working with 2D/3D animation programs, working on programming language, working with different software are only a few of the selections given to the students. Then they were given choices such as very often, often, sometimes, seldom and never to indicate how frequently they do this. The results are as follows: 65% connect to the internet, 38.34% listen to music and 5.84% work with different software very often. 37.50% examine educational materials, 37.08% study and do homework sometimes. They rarely use computers to develop projects (31.76%) or play games (22.10%). The rest (28.75%) never use computers to develop projects, don't work with any different software (31.76%), programming language (69.16%), or animation software (73.75%).

**Table 3.** For which purposes and how frequently do students use computers.

|  |  | Never | Seldom | Sometimes | Often | Very often | $\bar{X}$ | SD |
|---|---|---|---|---|---|---|---|---|
| Working with 2D/3D animation programs | F | 177 | 25 | 23 | 9 | 6 | | |
| | % | 73.75 | 10.42 | 9.58 | 3.75 | 2.50 | 1.51 | 0.98 |
| Working on programming language | F | 166 | 33 | 22 | 7 | 12 | | |
| | % | 69.16 | 13.75 | 9.17 | 2.92 | 5 | 1.61 | 1.09 |
| Working with different software | F | 92 | 57 | 57 | 20 | 14 | | |
| | % | 38.33 | 23.75 | 23.75 | 8.33 | 5.84 | 2.20 | 1.20 |
| Developing projects | F | 69 | 76 | 63 | 20 | 12 | | |
| | % | 28.75 | 31.67 | 26.25 | 8.33 | 5 | 2.29 | 1.12 |
| Playing games | F | 37 | 53 | 48 | 51 | 51 | | |
| | % | 15.40 | 22.10 | 20 | 21.25 | 21.25 | 3.11 | 1.37 |
| Examining educational materials | F | 22 | 42 | 90 | 54 | 32 | | |
| | % | 9.17 | 17.50 | 37.50 | 22.50 | 13.33 | 3.13 | 1.13 |
| Studying and doing homework | F | 4 | 24 | 89 | 86 | 37 | | |
| | % | 1.67 | 10 | 37.08 | 35.83 | 15.42 | 3.53 | 0.92 |
| Watching movies | F | 12 | 30 | 42 | 72 | 84 | | |
| | % | 5 | 12.50 | 17.50 | 30 | 35 | 3.78 | 1.19 |
| Listening to music | F | 20 | 32 | 39 | 57 | 92 | | |
| | % | 8.33 | 13.33 | 16.25 | 23.75 | 38.34 | 3.70 | 1.32 |
| Connecting to the Internet | F | 2 | 6 | 18 | 58 | 156 | | |
| | % | 0.83 | 2.50 | 7.50 | 24.17 | 65 | 4.50 | 0.80 |

When we examine the means of students' answers, it is seen that working with 2D and 3D animation software ($\bar{X}$=1.51) and programming languages ($\bar{X}$=1.61) have the lowest means. Connecting to the Internet ($\bar{X}$=4.5) has the highest mean. In this sense, it is seen that students use the Internet for different purposes rather than developing their computer skills.

**Table 4.** Participants' views on using of information technologies according to different aspects.

| | Getting Information | Research Analysis | Game and Entertainment |
|---|---|---|---|
| Mean | 2.67% | 3.30% | 3.53% |

When we categorize items under three aspects as research, acquisition of information and games-entertainment, it can be seen that games-entertainment category ($\bar{X}$=3.53) has the highest mean (Table 4). The lowest mean belongs to the category of using computers to acquire information ($\bar{X}$=2.67).

Pearson Chi-Square significance value ($\chi^2$) was 0.440 for font preference between genders. According to this result, there isn't a statistically significant difference among most preferred fonts by students according to gender. The font style preference between genders was $\chi^2=0.140$, this means that there isn't a statistically significant difference among the most preferred font styles according gender. The font preference based on computer education was $\chi^2=0.825$, so there isn't a statistically significant difference in the distribution of the most preferred fonts according to whether students had computer education or not. The font style preference between whether students had computer education or not was $\chi^2=0.438$, so there isn't a statistically significant difference in the distribution of font style preference according to whether students had computer education or not. The font preference between having a computer or not was $\chi^2=0.468$, this means, there isn't a statistically significant difference in the distribution of the most preferred fonts according to whether students had a computer or tablet or not. The style preference between having a computer or not was $\chi^2=0.414$, this means there isn't a statistically significant difference in the distribution of most preferred styles according to whether students had a computer, tablet or not.

**Participants' personality types**

In this sub-section, we analyze our data according to personality types of human temperament. Computations were done in MS Excel 2007 (Liengme, 2009) and where necessary SPSS (Griffith, 2010) was used.

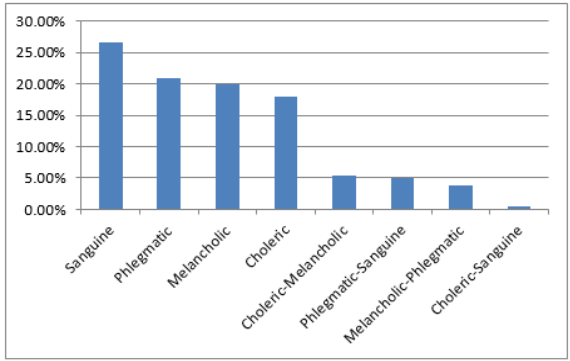 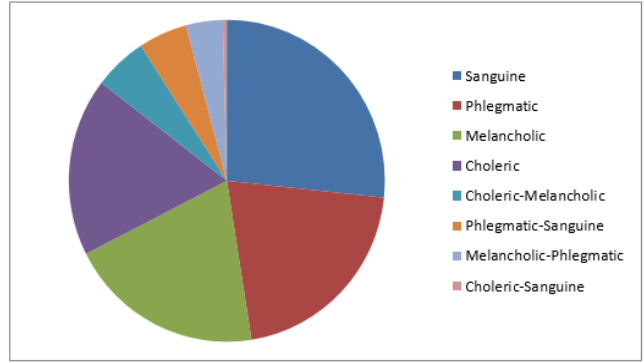

**Fig. 2.** Temperament types in percentage.   **Fig. 3.** Temperament types in pie chart.

The number of participants in this study is 240. After completing a survey partly based on Eysenck Personality Questionnaire (EPQ), the results are shown in Fig. 2 and Fig. 3. Looking at Fig. 2, 27% percent of the participants have a sanguine type of personality which is obviously the highest of them all. While on the other hand, a choleric-sanguine type of personality has the fewest students (1% only). The frequencies of other personality types are shown in Table 5.

| Table 5. Temperament Types Overall Result. | | | | | | | | |
|---|---|---|---|---|---|---|---|---|
| Sanguine | Phlegmatic | Melancholic | Choleric | Choleric-Melancholic | Phlegmatic-Sanguine | Melancholic-Phlegmatic | Choleric-Sanguine | Sum |
| 64 | 50 | 48 | 43 | 13 | 12 | 9 | 1 | 240 |
| 26.67% | 20.83% | 20.00% | 17.92% | 5.42% | 5.00% | 3.75% | 0.42% | 100% |
| Grade Point Average (GPA) | | | | | | | | Average |
| 83.63 | 82.55 | 85.36 | 80.79 | 83.08 | 83.78 | 81.63 | 78.00 | 83.10 |

Moreover, mixed types of temperaments (i.e. choleric-melancholic, phlegmatic-sanguine, melancholic-phlegmatic and choleric-sanguine) are fewer compared with the main type of temperaments (i.e. sanguine, phlegmatic, melancholic and choleric) as illustrated in Fig. 2. Therefore, there are only a few of the participants who fall into the mixed types of temperaments. According to our analysis of the 240 participants, 52.50% are emotionally stable, 43.33% are emotionally unstable, while the rest (4.17%) are neutral. The grade point average (GPA) of all the students in relation to temperament is shown in Table 5. Participants' GPA ranges from 67 to 94 (mean = 83.28, SD= 5.31) and 88% of the students have a GPA of at least 80.

While Li et al. et al. (2007) was defining sanguine, phlegmatic and phlegmatic-sanguine students as superior types in studying mathematics, our research shows that in general education all the personality types have approximately equal academic achievements[1].

---

[1] Here, we do not take into account the only student with choleric-sanguine personality type.

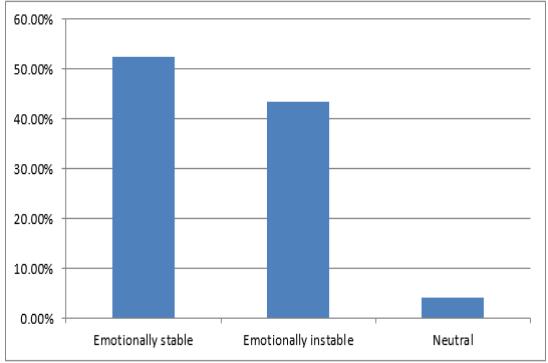 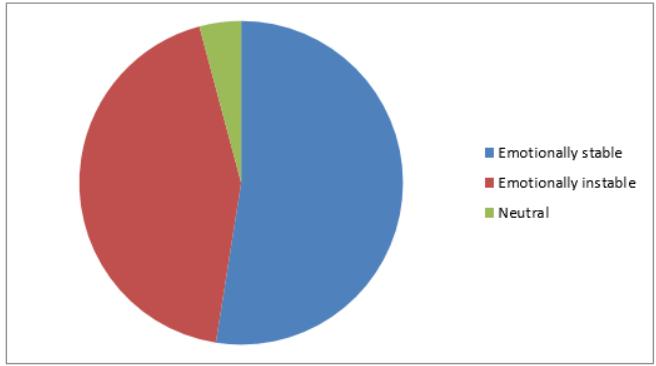

**Fig. 4.** Emotional stability.   **Fig. 5.** Emotional stability as a pie chart.

With the findings of emotional stability, the extraversion and introversion of participants were studied. These terms are clearly discussed in the third section of this manuscript. Looking at data gathered (Figs. 6-7) the percentage of extraverts (45.50%) and introverts (43.33%) slightly differ. 4.17% of the studied students are found to be neutral.

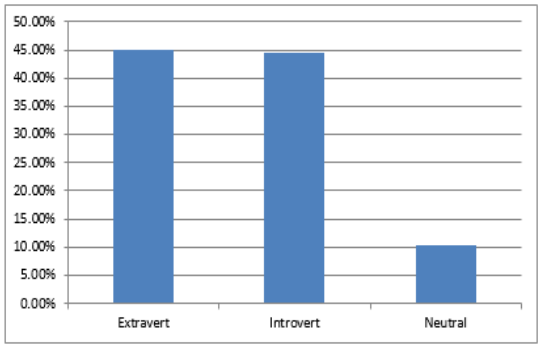 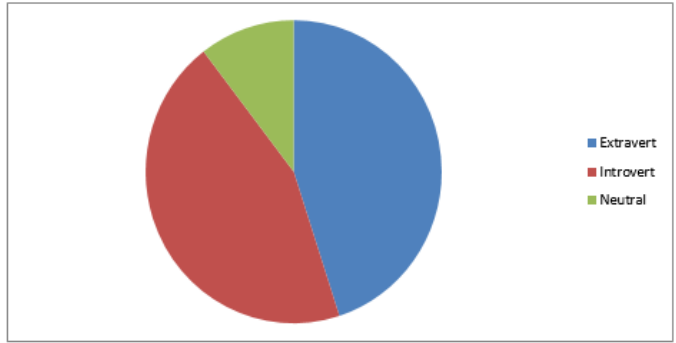

**Fig. 6.** Extraversion/Introversion.   **Fig. 7.** Extraversion/Introversion as a pie chart.

**Font and emphasis comprehension**

In this sub-section, the font and emphasis comprehension result is studied. The results were obtained from a survey partly based on the students' choice of favourite fonts (here we only focus on some of popular fonts) and emphasis on the text, and the results are shown in Table 6 and Figs 8-9. Fonts, as is commonly known, are divided into two main families: sans serif and sans. The results shown in Table 7 illustrate that sans serif and sans slightly differ.

| Table 6. Favourite font selection. | | | |
|---|---|---|---|
| Font | Font Family | Frequency | % |
| Times New Roman | Sans serif | 69 | 28.87% |
| Calibri | Sans | 34 | 14.23% |
| Cambria | Sans | 30 | 12.55% |
| Tahoma | Sans | 21 | 8.79% |
| Bookman Old Style | Sans serif | 21 | 8.79% |
| Gill Sans MT | Sans | 18 | 7.53% |
| Garamond | Sans serif | 16 | 6.69% |
| Verdana | Sans | 15 | 6.28% |
| Baskerville Old Face | Sans serif | 12 | 5.02% |
| Batang | Sans serif | 3 | 1.26% |
| Total | | 239 | 100.00% |

Among the given choices of fonts in the survey, Times New Roman, Calibri and Cambria are the three most frequently fonts. Times New Roman, belonging to Sans serif family, received 28.87% while Calibri and Cambria, both from Sans families received 14.23% and 12.55% respectively. Our result has shown that Sans serif family fonts have a total of 50.63% percentage of choice, and the Sans family has 49.37%. In conclusion, Sans serif and Sans are used almost equally in our research.

The emphasis of texts is sub-divided into bold, italic, highlight and underline. The results have shown that out of 240 students, a majority of them favoured bold with 62.50%, while italic 17.92% and highlight 14.17% were less favoured. Underline accounts for 5.42%. Figs. 8-9 show the results of both font choice and text emphasis.

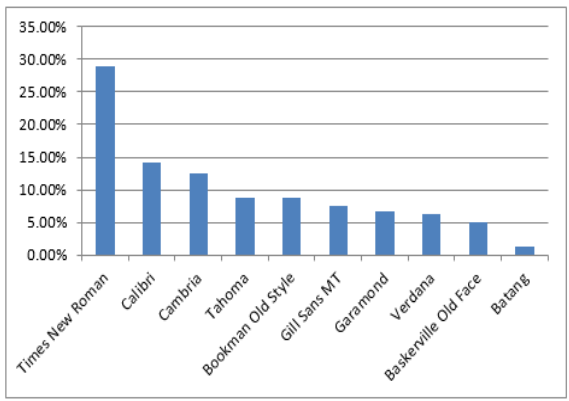
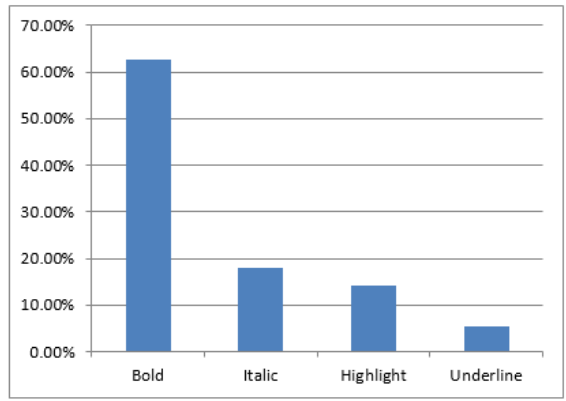

**Fig. 8.** Font choice.                **Fig. 9.** Text emphasis.

**Colours and colour schemes**

**Colours**

The paper continues to look into the subjects' (students') temperament with the corresponding colours they chose as part of the survey's questions. Students' were given the chance of naming their two favourite colours. In this study, colour selection or groupings were divided into three: warm, cool and achromatic colours. Warm colours are red, yellow, pink, maroon, orange, light red and golden yellow. Cool colours on the other hand are blue, green, violet, navy blue, turquoise, light green, ice blue, dark navy blue, aquamarine and magenta. Black, white, gray, cream and dark gray are achromatic colours. The study did not limit them on their choice of colours. The results have shown 20.44% of all the 240 students favoured blue, followed by black (16.48%), green (10.77%) and so on. Table 7 shows all results gathered.

**Table 7.** Students' favourite colours.

| Colours | Frequency | % |
|---|---|---|
| Blue | 93 | 20.44% |
| Black | 75 | 16.48% |
| Green | 49 | 10.77% |
| Red | 44 | 9.67% |
| Violet | 34 | 7.47% |
| White | 30 | 6.59% |
| Yellow | 30 | 6.59% |
| Navy blue | 25 | 5.49% |
| Turquoise | 20 | 4.40% |
| Pink | 15 | 3.30% |
| Maroon | 14 | 3.08% |
| Gray | 9 | 1.98% |
| Light green | 3 | 0.66% |
| Ice blue | 2 | 0.44% |
| Orange | 2 | 0.44% |
| Light blue | 2 | 0.44% |
| Light red | 2 | 0.44% |
| Cream | 1 | 0.22% |
| Dark navy blue | 1 | 0.22% |
| Aquamarine | 1 | 0.22% |
| Golden yellow | 1 | 0.22% |
| Magenta | 1 | 0.22% |
| Dark gray | 1 | 0.22% |
| Total | 455 | 100.00% |

Overall results shown in Table 7 illustrate the percentage of warm (23.47%) and achromatic (25.49%) colours slightly differ. The sum of warm and achromatic colours was almost equal to the percentage of cool colours (50.77%). Fig. 10 gives another view of the obtained results.

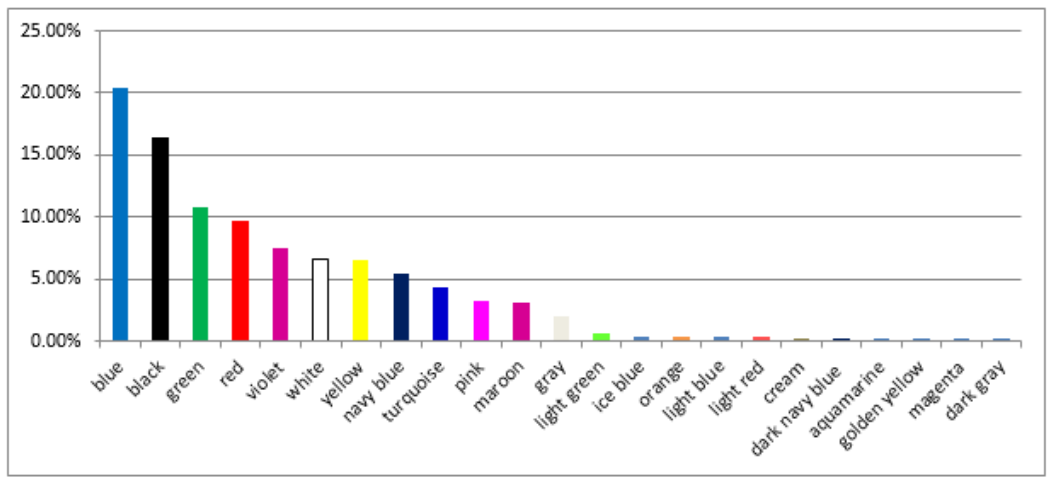

**Fig. 10.** Overall colour selection.

**Colours and personality types**

Additionally, the study compared the percentage of cool, warm and achromatic colours with the respective temperaments. Table 8 shows how each personality type has chosen the colour groups.

**Table 8.** Personality types with colour groups.

| Personality type | Cool colours | Warm colours | Achromatic | Total |
|---|---|---|---|---|
| Sanguine | 49.14% | 24.14% | 26.72% | 100.00% |
| Phlegmatic | 53.06% | 22.45% | 24.49% | 100.00% |
| Melancholic | 54.95% | 20.88% | 24.18% | 100.00% |
| Choleric | 44.71% | 28.24% | 27.06% | 100.00% |
| Choleric-Melancholic | **39.31%** | **30.43%** | **30.43%** | 100.00% |
| Phlegmatic-Sanguine | **60.87%** | **17.39%** | **21.74%** | 100.00% |
| Melancholic-Phlegmatic | 52.94% | 23.53% | 23.53% | 100.00% |
| Choleric-Sanguine[2] | Not computed | Not computer | Not computed | Not computed |

Analyzing the maximum and minimum percentage of colour groups pertaining to the personality types has shown the following results: for cool colours, the maximum percentage is 30.43% (choleric-melancholic) and the minimum percentage is 17.39% (phlegmatic-sanguine). For warm colours the maximum percentage is 60.87% (phlegmatic-sanguine) while the minimum percentage is 39.31% (choleric-melancholic). The maximum percentage in achromatic colours is 30.43% (choleric-melancholic) and the minimum percentage is 21.74% (phlegmatic-sanguine). Based on the above results, both choleric-melancholic and phlegmatic-sanguine gained the lowest and highest percentages in different colour groups.

Tables 9-15 show the results of different personality types based on colour selection. We restrict the result by frequency >5%.

**Table 9.** Sanguine.

| Colours | Blue | Black | Green | Red | White | Yellow | Violet | Turquoise |
|---|---|---|---|---|---|---|---|---|
| Percentage | 19.01% | 14.88% | 13.22% | 10.74% | 8.26% | 8.26% | 5.79% | 5.79% |

**Table 10.** Phlegmatic.

| Colours | Blue | Black | Green | Red | White | Yellow | Navy blue | Turquoise | Violet |
|---|---|---|---|---|---|---|---|---|---|
| Percentage | 21.43% | 16.33% | 10.20% | 9.18% | 7.14% | 7.14% | 7.14% | 7.14% | 5.10% |

**Table 11.** Melancholic.

| Colours | Blue | Black | Violet | Green | Red | Yellow | Navy blue |
|---|---|---|---|---|---|---|---|
| Percentage | 27.47% | 17.58% | 10.99% | 9.89% | 8.79% | 5.49% | 5.49% |

**Table 12.** Choleric.

| Colours | Black | Blue | Violet | Red | Green | Yellow | Pink |
|---|---|---|---|---|---|---|---|
| Percentage | 20.73% | 14.63% | 10.98% | 8.54% | 7.32% | 7.32% | 6.10% |

**Table 13.** Choleric-Melancholic.

| Colours | Black | Green | Maroon | Blue | Red | White | Pink |
|---|---|---|---|---|---|---|---|
| Percentage | 17.39% | 13.04% | 13.04% | 8.70% | 8.70% | 8.70% | 8.70% |

**Table 14.** Phlegmatic-Sanguine.

| Colours | Blue | Green | Red | White | Navy blue | Black |
|---|---|---|---|---|---|---|

---

[2] Since we had only student with this personality type, we could not compute corresponding cells.

| Percentage | 30.43% | 13.04% | 13.04% | 13.04% | 13.04% | 8.70% |
|---|---|---|---|---|---|---|

**Table 15.** Melancholic-Phlegmatic.

| Colours | Blue | Black | Green | Red | Turquoise | Violet | White | Yellow | Navy blue | Pink | Gray |
|---|---|---|---|---|---|---|---|---|---|---|---|
| Percentage | 17.65% | 11.76% | 11.76% | 11.76% | 11.76% | 5.88% | 5.88% | 5.88% | 5.88% | 5.88% | 5.88% |

The results have shown that, blue is the most chosen colour by most of the personality types followed by black and green.

**Colours schemes**

After the tested students wrote their two favourite colours, we analyzed them based on different colour schemes, namely: complement, triadic, analogous and split-complementary colour schemes (Williams, 2012). The tested students selected 23 different colours in total. This means, that following well-known rules of combinatorics (Brualdi, 2010) we can consider

$$C_{23}^2 = \frac{23!}{2!(23-2)!} = 253$$

complementary colour schemes. Using the same formula for computing the number of combinations we get the number of complementary, triadic, analogous and split-complementary colour schemes:

$$C_{23}^3 = \frac{23!}{3!(23-3)!} = 3542.$$

Our results have shown that for sanguine students who like blue and green, red can be added, in order to achieve a triadic colour scheme and visual harmony. Out of the 3542 possible combinations computed in CAS Maple for the triadic, analogous and split-complementary colour schemes our data have shown only one combination (i.e. blue, green, and red) present in the sanguine personality type. For the other personality types, none of the mentioned colour schemes exist. For two colours, on the other hand, with 253 possible combinations, our data have shown the same complementary colours – blue and yellow, green and violet. In regard to mixed temperaments, complementary colour schemes were not found.

**Differences in selecting colours by personality types**

The correlation between different personalities' colour selection is shown in Table 16. According to our research there are personality types which differ with colour choice such as choleric-sanguine and phlegmatic-sanguine. The correlation between them is 0.51. Moreover, a similar situation exists between melancholic and choleric-melancholic (0.62), sanguine and choleric-melancholic (0.69), and there are also some which correlate almost as a whole. For instance, the correlation between sanguine and phlegmatic colour selection is 0.93, phlegmatic and melancholic colour selection is 0.93 and so on. This shows that colour choice similarity between them is high. Table 16 shows more detailed results.

**Table 16.** Correlation matrix which shows how similarly colours were selected by different personality types.

| Personality Types | Sanguine | Phlegmatic | Melancholic | Choleric | Choleric-Melancholic | Phlegmatic-Sanguine | Melancholic-Phlegmatic |
|---|---|---|---|---|---|---|---|
| Sanguine | | 0.93 | 0.89 | 0.87 | **0.69** | 0.82 | 0.89 |
| Phlegmatic | | | 0.93 | 0.87 | **0.67** | 0.88 | 0.81 |
| Melancholic | | | | 0.89 | **0.62** | 0.87 | 0.81 |

|  |  |  |  |  |
|---|---|---|---|---|
| Choleric |  | 0.79 | **0.65** | 0.82 |
| Choleric-Melancholic |  |  | **0.51** | **0.66** |
| Phlegmatic-Sanguine |  |  |  | 0.79 |
| Melancholic-Phlegmatic |  |  |  |  |
| Choleric-Sanguine |  |  |  |  |

It is noted that since choleric-sanguine has only one student, necessary computations and correlations were not computed.

**Geometric Shapes**

Vector shapes created by means of B-splines and Bernstein-Bézier curves are one of the most important tools in computer animation. They allow us to create objects with different topologies, and are mostly studied in areas like computer aided geometric design (CAGD) (Farin, 2002) and computational mathematics. However, for creating animations with impressive characters, generating visually pleasing shapes plays an important role. The research work of Nabiyev & Ziatdinov (2014) notes that planar quadratic Bernstein-Bézier curves with monotonic curvature function (known as an aesthetic curve) may not be aesthetic, and in this sub-section we have empirically confirmed this. Tested students were selecting five visually pleasing ones among 24 Bernstein-Bézier curve segments of different topology, and the overall results of this part of our survey are shown in Fig. 11.

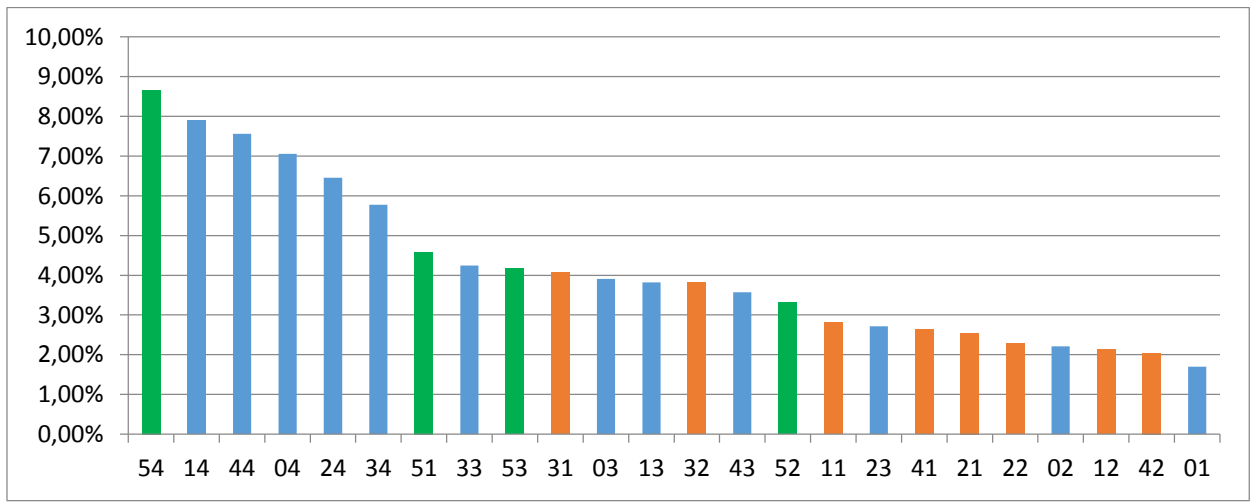

**Fig. 11.** Overall shape selection (numbers under polygons mean the name of jpg file and are not connected to any values here).

In our survey, students were asked to choose five curve segments out of 24. All planar curves were divided into three different types: MC-curves (Ziatdinov et al., 2013), curves with central symmetry, and other Bernstein-Bézier curves. The colours of the graph represent the three curves respectively: red for the MC-curves, green for curves with central symmetry and blue for other curves. The results in Fig. 11 have shown polygon 54, belonging to the curves with central symmetry, has the highest percentage of 8.67%. Polygon 14 with other curves followed having 7.90%. While polygon 31 belonging to MC-curve gathered only 4%. With an overall computation, MC-curves were selected 22%, while curves with central symmetry, 21%, and other curves were selected with 57%. Here, we can conclude that monotonicity of a curvature function as in MC-curves does not guarantee that a curve segment is visually pleasing.

However, the situation may be a bit different, if we differentiate among personality types. Figures 12-19 and Table 17 show these cases.

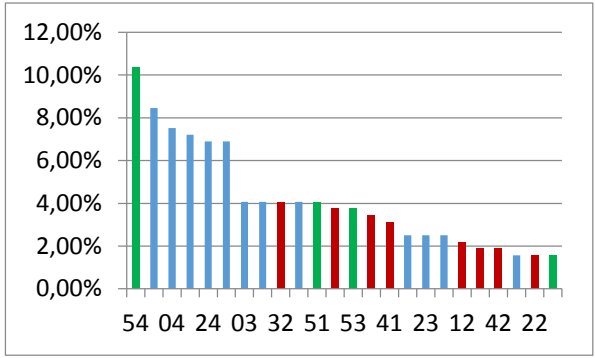

**Fig. 12.** Sanguine.

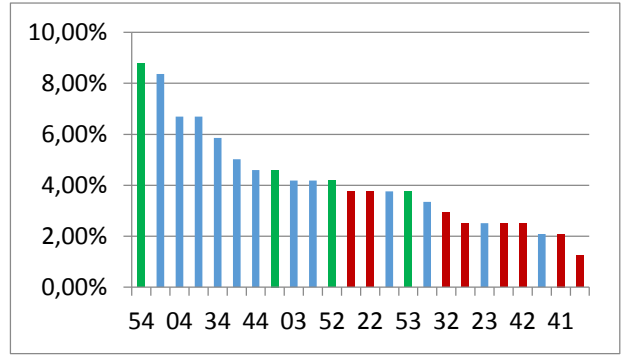

**Fig. 13.** Phlegmatic.

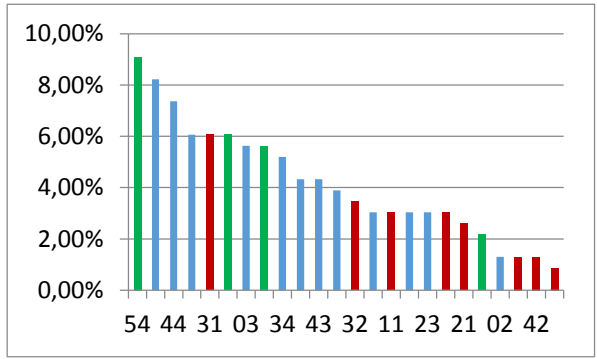

**Fig. 14.** Melancholic.

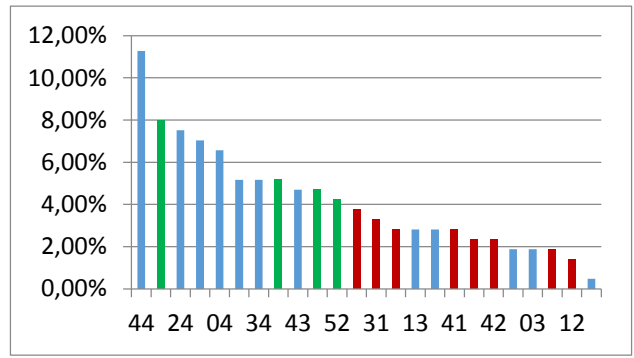

**Fig. 15.** Choleric.

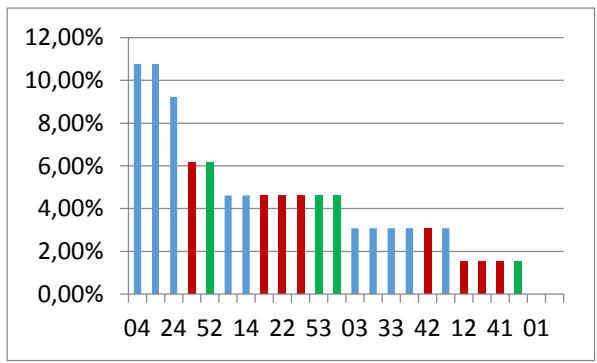

**Fig. 16.** Choleric-melancholic.

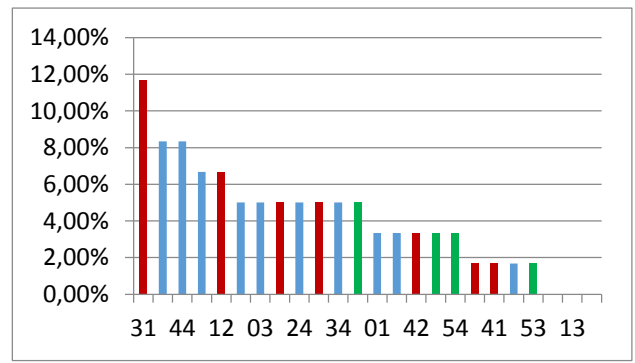

**Fig. 17.** Phlegmatic-sanguine.

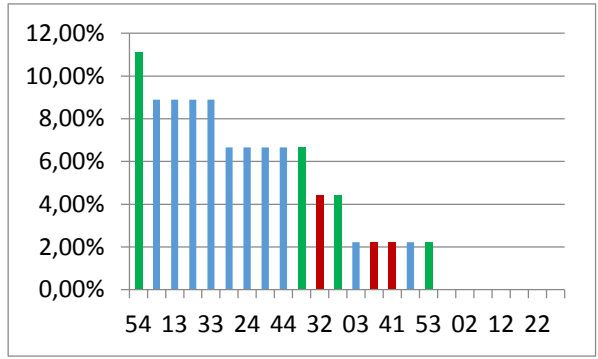 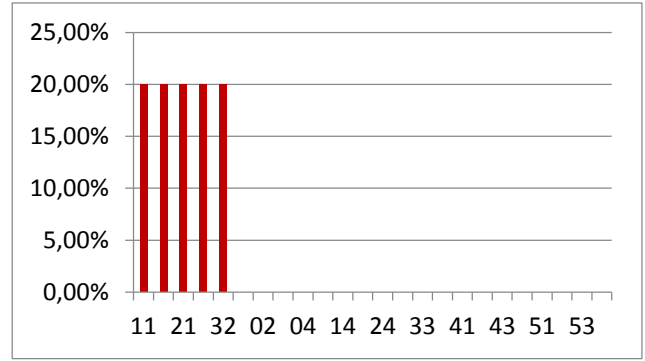

**Fig. 18.** Melancholic-phlegmatic.   **Fig. 19.** Choleric-sanguine.

**Table 17.** How curve segments were selected by different personality types.

| Personality Types | MC-curves | Curves with central symmetry | Other curves |
|---|---|---|---|
| Sanguine | 22% | 20% | 58% |
| Phlegmatic | 23% | 23% | 54% |
| Melancholic | 27% | 24% | 49% |
| Choleric | 11% | **11%** | **78%** |
| Choleric-Melancholic | **47%** | **29%** | **24%** |
| Phlegmatic-Sanguine | 37% | 14% | 49% |
| Melancholic-Phlegmatic | **10%** | 28% | 63% |
| Average | 22% | 21% | 57% |

It is noted that since choleric-sanguine has only one student, necessary computations were not carried out.

In Table 17 it is shown that the difference between the frequency of MC-curves selected by choleric (11%) and melancholic-phlegmatic (10%) is only one percent and it's not significant. Obtained results suggest that choleric-melancholic students are attracted most by MC-curves and curves with central symmetry among all the personality types, but choleric students show opposite results. The cells filled with red are the maximum frequency of the selected curve. While those which are filled with blue are the minimum.

**Computer animation principles**

This sub-section analyses the results gathered from the last part of the survey regarding the eleven principles of traditional computer animation (Lasseter, 1987): squash and stretch, timing, anticipation, staging, follow through and overlapping action, straight ahead action and pose-to-pose action, show in and out, arcs, exaggeration, secondary action and appeal. Students were asked eleven multiple type questions, and the following results were obtained.

**Table 18.** How animation principles were answered by different personality types.

| Question No. | Personality types | | | | | | | | Average by question |
|---|---|---|---|---|---|---|---|---|---|
| | Sanguine | Phlegmatic | Melancholic | Choleric | Choleric-Melancholic | Phlegmatic-Sanguine | Melancholic-Phlegmatic | Choleric-Sanguine | |
| 1. | 30% | **14%** | 29% | 19% | 31% | 25% | **33%** | 0% | 25% |
| 2. | 66% | 59% | 74% | 60% | 77% | **83%** | **56%** | 0% | 66% |
| 3. | **44%** | **44%** | 23% | 30% | 31% | 27% | **11%** | 0% | 34% |

*16*

| | | | | | | | | | |
|---|---|---|---|---|---|---|---|---|---|
| 4. | 63% | 60% | 53% | 65% | **42%** | **67%** | **67%** | 0% | 60% |
| 5. | 38% | 26% | 29% | 30% | **46%** | **25%** | 33% | 100 | 32% |
| 6. | 75% | 70% | 69% | 72% | **77%** | 75% | **44%** | 0% | 71% |
| 7. | 56% | 58% | 48% | 60% | 46% | **67%** | **33%** | 0% | 54% |
| 8. | **54%** | **23%** | 43% | 49% | 31% | 50% | 33% | 0% | 47% |
| 9. | 19% | 28% | 19% | 14% | **31%** | **8%** | 0% | 0% | 19% |
| 10. | 55% | **59%** | 52% | **37%** | 42% | 50% | 44% | 0% | 50% |
| 11. | 81% | 80% | 69% | **65%** | 69% | **92%** | 67% | 100 | 75% |
| Average | **52,82%** | 47,36% | 46,18% | 45,55% | 47,55% | 51,73% | **38,27%** | | 44,42% |
| Average for main and mixed personality types | | | | | | | | | |
| Main Temperaments | | | | Mixed Temperaments | | | | | |
| 47,97% | | | | 45,85% | | | | | |

Table 18 shows that the maximum and minimum frequencies of correctly selected questions correspond to mixed personality types (except for choleric-sanguine as it was mentioned in previous sections). At the age of 14-17 school students are not aware of the fundamental principles of computer animation, so obviously, these questions were answered by their own logical understanding. The overall average frequency of correct answers was 44,42%. The results shown in Table 18 of overall frequencies of correctly selected questions illustrate sanguine (52.82%) as the highest temperament type while melancholic-phlegmatic (38.27%) on the other hand has the least. Among all the studied personality types, melancholic-phlegmatic students have shown the least readiness in understanding the fundamental principles for creating computer animations, on the other hand sanguine and phlegmatic-sanguine students were the best. In general, main and mixed personality types were giving correct answers similarly. There are some aspects like exaggeration in computer animation, and only 19% of students were able to understand it.

## Limitations

- The results of our survey have shown that we had only one choleric-sanguine personality, so this mixed personality type was not studied in as much detail as others;
- We have not studied in detail how different personality types prefer an object to be moved on their screens, and what should be the topology of used paths. This would be a very important variable since animation is known as the art of movement.

We believe that these remarks should be taken into account in our future work or by other researchers who find them important to be considered in temperament-based learning.

## Conclusion and future work

The aim of the current work was to establish an inter-disciplinary field of research towards greater educational effectiveness where psychology, colour science, computer animation, geometric modelling and technical aesthetics intersect. Our research was based on a survey which was conducted among school students aged 14-17. For the first time in computer-assisted learning we have proposed a way for developing educational computer animation based on fundamental personality types of human temperament. Our experimental analysis has shown how fundamental principles of traditional computer animation are understood by school students. In addition, we have found that both choleric-melancholic and phlegmatic-sanguine students gained the lowest and highest percentages in selection of different groups of students' favourite colours (warm, cool, and achromatic). The students' choices of aesthetic shapes used in computer animation were used to experimentally confirm the theory of Nabiyev and Ziatdinov (2014) which reports that planar quadratic Bernstein-Bézier curves with monotonic curvature function may not be aesthetic. Finally, we have shown that choleric-melancholic students are attracted most by MC-curves and curves with central symmetry among all the personality types, but choleric students are attracted by other Bernstein-Bézier curves. The present study believes that it is likely to have wide benefits in the field of education. Considering the personality types of students' temperament in designing educational computer animations with the aid of gathered empirical results might be a promising avenue to enhance the learning process.

We envisage potential applications of our findings in learning graphic design. Our research has shown that different personality types perceive the shapes in different ways, and this can be used as a tool to communicate pictorial messages to the audience in a more efficient way. One of the other future directions can be studying relationships between personality types and their achievement in social sciences and arts. All these questions will be studied in our future manuscripts.

**Acknowledgement**

We would like to thank Dr. Rifkat I. Nabiyev (Ufa State Aviation Technical University, Ufa, Russia) for his valuable comments, and Huseyin Kinay (Fatih University, Istanbul, Turkey) for his assistance during this research. The first author of this work was supported by the Scientific Research Fund of Fatih University under the project number P55011301_Y (3141).
The second author is very grateful to Prof. Kenjiro T. Miura (Shizuoka University, Hamamatsu, Japan) and his Realistic Modelling Lab. members for discussion on monotone-curvature curves.